\documentclass[sigconf,nonacm]{acmart}

\usepackage{booktabs}
\usepackage{multirow}
\usepackage{graphicx}
\usepackage{amsmath}
\usepackage{xspace}
\usepackage{cleveref}

\newcommand{\dataset}{MultiMental-IAS\xspace}
\newcommand{\datasetdoi}{10.18419/DARUS-XXXX}

\newif\ifvalidated
\validatedfalse

\begin{document}

\title{MultiMental-IAS: A Multimodal Driver Mental State Analysis\\Dataset for Context-Aware Non-IID Evaluation}

\author{Baran Can G\"ul}
\affiliation{%
  \institution{Institute of Industrial Automation and Software Engineering (IAS), University of Stuttgart}
  \city{Stuttgart}
  \country{Germany}}
\email{baran-can.guel@ias.uni-stuttgart.de}

\author{Nagendra Pal Singh}
\affiliation{%
  \institution{Institute of Industrial Automation and Software Engineering (IAS), University of Stuttgart}
  \city{Stuttgart}
  \country{Germany}}
\email{st191659@stud.uni-stuttgart.de}

\author{Anjana Arvind Naik}
\affiliation{%
  \institution{Institute of Industrial Automation and Software Engineering (IAS), University of Stuttgart}
  \city{Stuttgart}
  \country{Germany}}
\email{anjananaik04@gmail.com}

\renewcommand{\shortauthors}{G\"ul et al.}

\begin{abstract}
Robust driver-state modelling requires synchronized physiological, behavioural, and contextual signals, yet existing open resources typically target a single cognitive state, omit continuous vehicle context, or do not define evaluation protocols that expose inter-subject heterogeneity. We present \dataset, a multimodal driver mental-state dataset collected in a cockpit-in-the-loop simulation built on CARLA\@. It contains vehicle telemetry with scenario-context channels, four-channel EEG and three-channel PPG from a Muse~2 headband, and cabin-facing video from 23 participants, recorded in 148 trials (11.6~hours) under six induced states: Normal, Distracted, Alerted, Stressed, Relaxed, and Drowsy. We document the acquisition pipeline in detail and provide a benchmark with one protocol for all models: chronological within-subject splits, participant-disjoint cross-subject folds, validation-based model selection, and ablations over modalities, scenario-context channels, and signal normalization.
Multimodal fusion reaches a macro-F1 of 0.94 for drivers seen during training. For unseen drivers, fusing driving dynamics with per-window normalized EEG/PPG raises macro-F1 from 0.408 (telemetry alone) to 0.485, consistently across all five folds, while the gap between the two protocols quantifies the inter-driver heterogeneity that personalized models must address.
\dataset thus provides a concrete testbed for non-IID generalization, personalization, and federated learning in driver monitoring.
\end{abstract}


\maketitle

\section{Introduction}
Transitions between automation levels in semi-autonomous and software-defined vehicles require the human driver to maintain or promptly resume cognitive engagement~\cite{stumpfle2025}. Fatigue, distraction, stress, and drowsiness each impair reaction time and situational awareness in distinct ways, and their signatures overlap considerably when individual sensor streams are considered in isolation~\cite{rivasvidal2026}. Combining vehicle dynamics, physiological measurements, and cabin-facing video offers a more complete picture of driver state, and multimodal learning has consequently attracted considerable attention in automotive human-centred computing~\cite{sikander2019}.

Publicly available datasets, however, still present three recurring limitations. First, most resources focus on a single target state, typically drowsiness or vigilance, and do not capture the broader spectrum of overlapping cognitive conditions relevant to driving safety~\cite{zheng2017, massoz2016drozy}. Second, few datasets synchronize physiological streams with continuous vehicle context channels such as traffic density and illumination, even though these factors modulate arousal and workload~\cite{distracted2017}. Third, as summarized in Table~\ref{tab:comparison}, existing datasets rarely define evaluation protocols that separate within-subject from cross-subject performance, which hinders assessment of generalization across users.

This third limitation is particularly consequential in safety-critical deployments. Physiological baselines vary substantially across individuals: resting cortical rhythms, baseline heart-rate variability (HRV), and the mapping between mental state and EEG spectral power differ with age, health status, and experience~\cite{saha2020eeg}. Models optimized on pooled cohorts frequently degrade on unseen users, a problem formalized in the federated learning literature as non-IID data~\cite{li2020federated, karimireddy2020}. Addressing this gap requires resources that treat inter-subject heterogeneity as a first-class property rather than an uncontrolled nuisance.

\begin{table*}[t]
\centering
\caption{Comparison with representative open datasets in multimodal driver monitoring and physiological affective computing. Modalities: E = EEG/EOG, P = PPG/ECG/EDA/EMG, V = video, T = telemetry. $^{a}$Face video for 22 of the 32 participants.}
\label{tab:comparison}
\small
\setlength{\tabcolsep}{4pt}
\renewcommand{\arraystretch}{1.12}
\begin{tabular}{lrrllll}
\toprule
Dataset & Year & Subj. & Modalities & Primary Labels & Acquisition setting & States \\
\midrule
MAHNOB-HCI \cite{soleymani2012mahnob} & 2012 & 27 & E, P, V & Affective response & Lab (video stimuli) & 5 \\ 
DEAP \cite{koelstra2012deap} & 2012 & 32 & E, P, V$^{a}$ & Valence/arousal/dominance/liking & Lab (music-video stimuli) & cont. \\
DROZY \cite{massoz2016drozy} & 2016 & 14 & E, P, V & Drowsiness / KSS & Lab (PVT, sleep deprivation) & cont. \\
SEED-VIG \cite{zheng2017} & 2017 & 23 & E & Vigilance / PERCLOS & Driving simulator & cont. \\
Distracted Driving \cite{distracted2017} & 2017 & 68 & P, V, T & Distraction condition & Driving simulator & 4 \\
AffectiveROAD \cite{haouij2018affectiveroad} & 2018 & 10 & P, V & Stress / attention & Real-world urban driving & cont. \\
DD-Database \cite{dddatabase2023} & 2023 & 10 & E, P & Drowsiness events & Driving simulator & 1 \\
MPDB \cite{mpdb2024} & 2024 & 35 & E, P, T & Driving behaviour & Driving simulator & 5 \\
\midrule
\textbf{\dataset (ours)} & \textbf{2026} & \textbf{23} & \textbf{E, P, V, T} & \textbf{Driver mental states} & \textbf{CARLA simulator + cabin} & \textbf{6} \\
\bottomrule
\end{tabular}
\end{table*}

We introduce \dataset, a cockpit-in-the-loop dataset for six-class driver mental-state analysis. Among the resources in Table~\ref{tab:comparison}, it is the only one that combines vehicle telemetry (including scenario context channels), EEG, PPG, and video across six mental states. Our contributions are:
\begin{itemize}
  \item A four-modality dataset (telemetry, EEG, PPG, and cabin-facing video) of 148 trials from 23 participants across six induced driver mental states, released with the original host timestamps, corrected channel labels, and per-trial metadata.
  \item A detailed account of the acquisition and preprocessing pipeline, with per-trial metadata (durations, effective sampling rates, scenario settings) that lets users select and preprocess the data for their own analyses.
  \item A reproducible benchmark with one protocol for all models: a chronological within-subject split with a held-out validation segment, participant-disjoint cross-subject folds, validation-based model selection, and ablations over modalities, scenario-context channels, and signal normalization.
  \item Reference results showing that multimodal fusion reaches a macro-F1 of 0.94 within subject and that fusing driving dynamics with normalized physiology improves recognition for unseen drivers (0.485 vs.\ 0.408, better in all five folds).
\end{itemize}

\section{Related Work}
Research on inferring driver cognitive state from physiological and behavioural signals spans three decades, yet gaps in modality coverage, contextual sensing, and cross-subject benchmarking persist~\cite{sikander2019, baltrusaitis2019multimodal}.

\subsection{Physiological and Behavioural Biomarkers}
EEG spectral dynamics provide the most direct cortical measure of driver state. Alpha-band power (8--13~Hz) is commonly associated with reduced arousal and tends to increase during fatigue, while task-related cognitive load is often accompanied by alpha suppression. Frontal theta activity (4--8~Hz), in contrast, has been linked to sustained mental effort and increases with workload~\cite{wascher2023eeg, getzmann2021workload}.

Cardiac signals offer a complementary autonomic perspective: RMSSD and the LF/HF ratio of HRV respond to acute stress and are used as markers of autonomic activation~\cite{lehrer2020heart, thayer2006beyond}. In-cabin video contributes behavioural cues (blink rate, eyelid closure, head pose) that are informative for drowsiness and distraction, while vehicle telemetry captures the consequences of internal state changes in observable control behaviour~\cite{distracted2017, abouelnaga2017distracted}. Because physiological signals reflect internal changes that can precede visible behavioural deterioration, the modalities are expected to carry complementary information~\cite{baltrusaitis2019multimodal}. Consumer-grade dry-electrode headbands such as the Muse~2 have been shown to capture workload-related changes in frontal and temporal band power~\cite{leclercq2022muse, dobrzynska2024eeg}.

\subsection{Driver Monitoring Datasets}
SEED-VIG~\cite{zheng2017} and DROZY~\cite{massoz2016drozy} are foundational EEG and physiology benchmarks for vigilance and drowsiness estimation, yet both target a single vigilance or drowsiness dimension. The Distracted Driving dataset~\cite{distracted2017} provides rich simulator telemetry and peripheral physiology across distraction conditions from 68 participants, but contains no EEG. AffectiveROAD~\cite{haouij2018affectiveroad} captures stress and attention during real-world urban driving with wearable sensors, but is limited to ten participants and a single continuous label dimension.
MAHNOB-HCI~\cite{soleymani2012mahnob} and DEAP~\cite{koelstra2012deap} are influential multimodal affect datasets, but are grounded in passive media viewing without driving context or vehicle telemetry. More recent simulator-based resources combine EEG with other physiological signals in driving: DD-Database~\cite{dddatabase2023} annotates drowsiness events only, and MPDB~\cite{mpdb2024} labels five driving manoeuvres rather than mental states. MultiFuser~\cite{multifuser2024} demonstrates multimodal transformer fusion for driver action recognition rather than internal cognitive state.

\subsection{Non-IID Generalization and Federated Learning}
Intra- and inter-subject variability in EEG induces covariate shift that impedes model transferability~\cite{saha2020eeg}, producing non-IID data distributions under which models trained on pooled cohorts can degrade markedly on unseen users~\cite{li2020federated}. Federated learning~\cite{mcmahan2017fedavg} enables collaborative training across heterogeneous clients without sharing raw data. Its feasibility in automotive contexts has been shown for comfort feature prediction and collaborative sensing in software-defined vehicles~\cite{gul2024person, gul2024fed}, and for real-time multimodal emotion recognition in embedded settings~\cite{gul2025fedmultiemo}. None of these efforts addressed mental-state classification from synchronized physiological, behavioural, and vehicle streams across multiple states, which \dataset is designed to support.

\section{The \dataset Dataset}

\subsection{Apparatus and Data Acquisition}
\label{sec:apparatus}
\begin{figure}[t]
  \centering
  \includegraphics[width=\columnwidth]{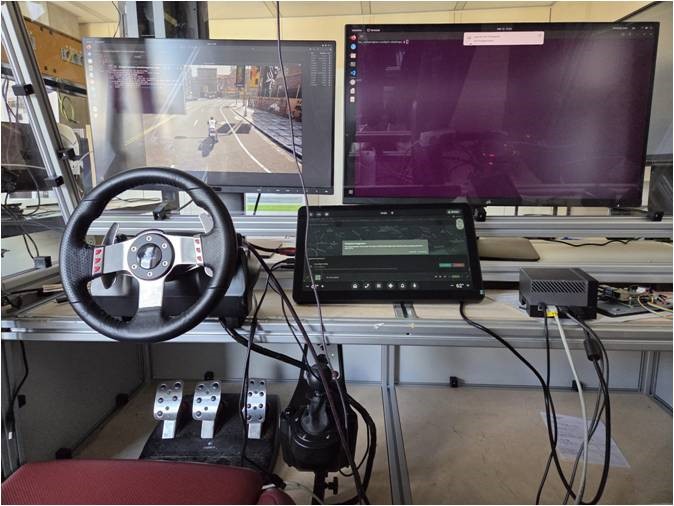}
  \caption{\dataset data-collection setup: IAS cockpit with Logitech G27 steering wheel and pedals, Muse~2 EEG/PPG headband, and Intel RealSense D455 cabin-facing camera.}
  \label{fig:cockpit}
\end{figure}

\begin{figure*}[t]
  \centering
  \includegraphics[width=0.94\textwidth]{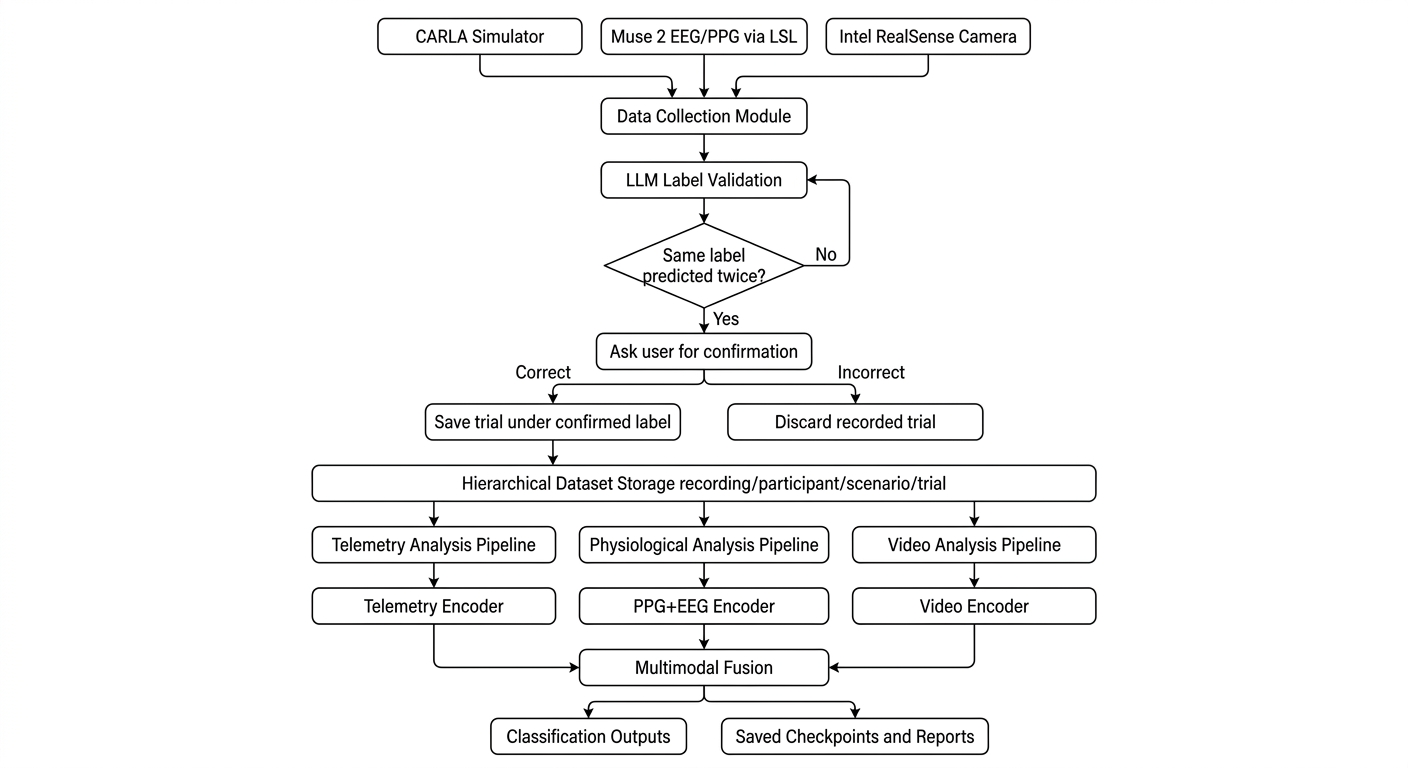}
  \caption{\dataset acquisition and processing pipeline: CARLA telemetry, Muse~2 EEG/PPG streams received via LSL, and Intel RealSense video are recorded per trial and stored hierarchically; parallel unimodal encoders and a fusion head form the reference model. The LLM-assisted confirmation loop is an optional acquisition aid (Section~\ref{sec:llm}).}
  \label{fig:system}
\end{figure*}

Data were collected in a cockpit-in-the-loop simulation running CARLA~0.9.13~\cite{dosovitskiy2017carla} on a workstation with an Intel Core i9-12900K processor, 64~GB RAM, and an NVIDIA RTX~3090 GPU\@. The cockpit used a Logitech~G27 force-feedback steering wheel and pedal set (Figure~\ref{fig:cockpit}). Figure~\ref{fig:system} shows the acquisition and processing pipeline.

Physiological signals were acquired with a Muse~2 headband, which provides four dry EEG channels (TP9, AF7, AF8, TP10; native rate 256~Hz), an auxiliary channel, and three optical PPG channels (PPG1--PPG3; native rate 64~Hz). The headband streamed over Bluetooth, and the recording host received the streams through the Lab Streaming Layer (LSL) using \texttt{pylsl}. Vehicle telemetry was obtained through the CARLA Python API and comprises position ($x, y, z$), orientation (yaw, pitch, roll), speed, 3-axis acceleration, and scenario context variables: traffic density, road type, sun altitude, time of day, ambient light, snow, and travelled distance. Cabin-facing video was recorded with an Intel RealSense D455 camera at $640 \times 480$ pixels; the camera shows the whole driver seat and part of the room behind it. 

\paragraph{Effective sampling rates:}
The recording logger wrote one row per stream in each acquisition cycle, timestamped with the host wall clock (millisecond resolution). The telemetry, EEG, and PPG logs of a trial therefore share one timestamp grid, and their effective rate is the rate of that cycle rather than the native sensor rate: the median over all trials is 20.3~Hz (range 9.7--30.0~Hz). In the \textit{Drowsy} condition, which is driven at night, the cycle ran at a median of 11.3~Hz (20.1--21.4~Hz for the other conditions), as rendering the night scene requires more computation. The logs therefore capture the slow components of the physiological signals, such as low-frequency EEG activity and the pulse waveform. The benchmark resamples all streams to a common band (Section~\ref{sec:preproc}), and effective rates are listed per trial in \texttt{metadata.csv}.

\subsection{Participants}
\label{sec:participants}
Twenty-three participants took part voluntarily (identifiers P11--P33; IDs P1--P7 belong to an earlier pilot that is not part of the release). %
Participants were aged 19--63 years ($M{=}30.5$, $SD{=}9.8$, median~27). Seven did not hold a driving licence; six reported corrected vision problems, two low blood pressure, and one diabetes. Participants with active cardiac arrhythmia, a pacemaker, self-reported photosensitive epilepsy, or severe simulator sickness were excluded at screening. Three of the authors took part as participants. All participants gave written informed consent. 

\subsection{Induction Protocol}
Each condition was recorded as a block of up to five minutes. States were induced through guided mental recall, behavioural imitation, explicit task instructions, pre-trial preparatory activities, and CARLA scenario settings. Specifically, \textit{Normal} was recorded under calm driving analogous to a daily commute; \textit{Distracted} involved simulated secondary-task glances and infotainment interactions; \textit{Alerted} required sustained hazard anticipation; \textit{Stressed} combined pre-trial stair-climbing, stressful music, and urgency instructions; \textit{Relaxed} used soothing music, recall of a peaceful memory, and minimal task demand; and \textit{Drowsy} recordings were scheduled when participants reported fatigue and included fatigue-related behavioural cues such as slow blinking and head nodding. 

Table~\ref{tab:states} summarizes the conditions together with the illumination recorded in the context channels. Five conditions were driven in daylight (sun altitude $45^\circ$); \textit{Drowsy} was driven at night in 28 of its 30 trials and at dusk in the other two. Because illumination and traffic settings are part of the induction, these context channels covary with the label by design.

\begin{table*}[t]
\caption{Conditions, logged illumination (from the context channels of all 148 trials), and hypothesized multimodal tendencies (expectations, not measured results).}
\label{tab:states}
\begin{tabular}{llllp{6.2cm}}
\toprule
State & Mental focus & Illumination (logged) & Trials & Hypothesized tendencies \\
\midrule
Normal     & Routine calm driving      & day (22), dusk (1)   & 23 & Stable speed, smooth steering, moderate physiological baseline \\
Distracted & Split attention           & day (22), dusk (2)   & 24 & Speed variability, lane deviation, gaze shifts, delayed braking \\
Alerted    & Sustained high vigilance  & day (20), dusk (2)   & 22 & Rapid corrections, heightened posture, suppressed frontal alpha \\
Stressed   & Urgency and arousal       & day (22), dusk (1)   & 23 & Harsh braking, aggressive steering, reduced HRV, elevated sympathetic tone \\
Relaxed    & Low arousal but attentive & day (24), dusk (2)   & 26 & Smooth low-variance control, elevated HRV, higher relative alpha \\
Drowsy     & Low vigilance / fatigue   & night (28), dusk (2) & 30 & Delayed corrections, inconsistent speed, physiological slowing \\
\bottomrule
\end{tabular}
\end{table*}

\section{Data Organization and Preprocessing}

\subsection{Release Structure}
\label{sec:organization}
\begin{figure}[t]
  \centering
  \includegraphics[width=\columnwidth]{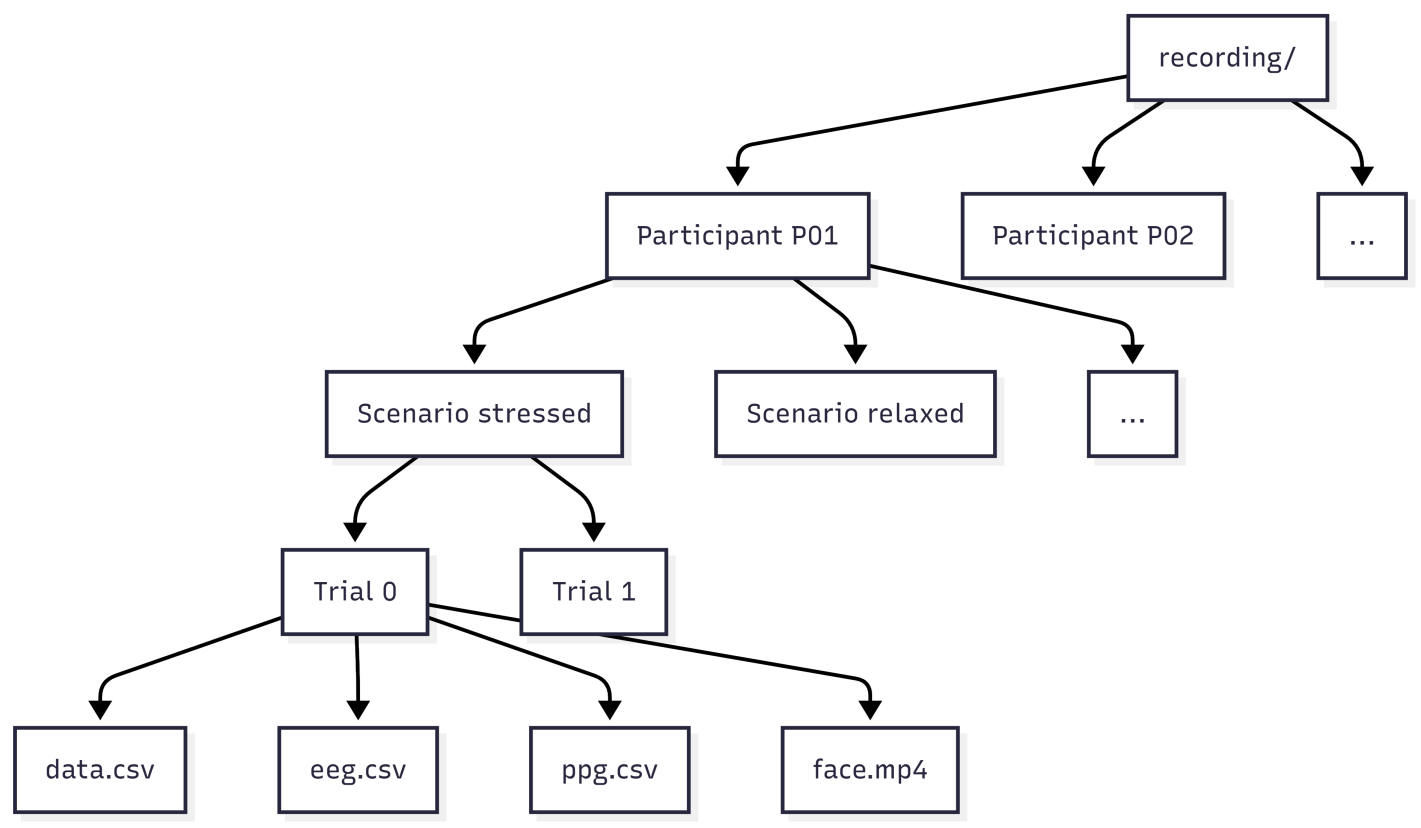}
  \caption{Hierarchical organization: one folder per participant, condition, and trial, each containing \texttt{data.csv}, \texttt{ppg.csv}, \texttt{eeg.csv}, and \texttt{face.mp4}.}
  \label{fig:dataset_org}
\end{figure}

The release is organized by participant, condition, and numbered trial (\texttt{\{participant\}/\{condition\}/\{trial\}/}), as shown in Figure~\ref{fig:dataset_org}. Each trial folder contains the telemetry log (\texttt{data.csv}), the PPG log (\texttt{ppg.csv}), the EEG log (\texttt{eeg.csv}), and the cabin video (\texttt{face.mp4}). Every CSV file starts with \texttt{time\_s}, the time since the first sample of the trial computed from the host timestamps, followed by the original timestamp (\texttt{timestamp\_\allowbreak local}) and the channels. A \texttt{metadata.csv} file lists, per trial, the start time, duration, row count, effective rate, largest gap, logged context settings, and video properties of every stream.

The release contains 148 trials (Table~\ref{tab:states}): 20 participants contributed all six conditions, and three contributed four or five. For several participants, additional recordings were collected in selected conditions to increase the number of samples; these additional trials vary in length, and 13 trials in total are shorter than five minutes (13--222~s). The total logged duration is 11.6~hours, and 146 of the 148 trials include video.

\subsection{Release Generation}
\label{sec:corrections}
The release is generated directly from the original logs by a provided script (\texttt{fix\_\allowbreak dataset.py}). It derives \texttt{time\_s} from the host timestamps, labels the EEG channels TP9, AF7, AF8, TP10, and AUX, and writes the per-trial metadata. Videos were normalized to a duration of 300~s during post-processing and are mapped linearly onto the logged duration of each trial.

\subsection{Benchmark Preprocessing}
\label{sec:preproc}
Each CSV stream is interpolated onto a uniform 20~Hz grid using \texttt{time\_s}, low-pass filtered with a zero-phase fourth-order Butterworth filter at 4~Hz, and decimated to 10~Hz. Because every trial, including the \textit{Drowsy} trials logged at about 11~Hz, then carries the same 0--4~Hz band, the sampling rate can no longer give away the condition. For EEG, this retains the delta band only. Telemetry is represented by speed, three-axis acceleration, yaw rate (from the unwrapped yaw angle), pitch, and roll; the scenario channels (traffic density, sun altitude, ambient light, snow) and the map position are used only in explicitly marked ablations. Videos are decoded at 5~fps, converted to grayscale, and downscaled to $48\times48$ pixels.

\subsection{LLM-Assisted Human-in-the-Loop Acquisition Tool}
\label{sec:llm}
The acquisition software includes an optional labelling aid. Every 10~s it samples four camera frames, detects the face with MediaPipe, and computes simple EEG, PPG, and telemetry statistics. It sends the frames and statistics to a locally hosted LLaVA-7B model (Ollama), which returns a label from the six conditions, a confidence score, and a short rationale. When the same label is returned for four consecutive windows (40~s), the operator is asked to accept it or to select a different label. The labels used in this paper are the induced condition of each trial, not LLM outputs.

\section{Benchmark}
\label{sec:benchmark}
The benchmark asks two questions: how well can each modality and combination recognize the induced state of a driver whose other recordings were seen during training, and how well does this transfer to drivers who were never seen? A single script (\texttt{train\_\allowbreak benchmark.py}) implements the protocol for all models.

\subsection{Protocol}
All models classify 10-s windows with a 5-s stride into the six conditions; each window inherits the label of its trial.
\textit{Within-subject:} in every trial, windows lying entirely in the first 70\% of the trial are used for training, those in the 70--85\% segment for validation, and those in the last 15\% for testing. Windows that cross a boundary are discarded, so no sample appears in two splits (5{,}548 / 962 / 962 windows).
\textit{Cross-subject:} participants are divided into five disjoint folds; in each fold, three of the training participants are held out for validation, and all windows of the test participants are evaluated (8{,}047 windows over the five folds).
Inputs are z-scored with statistics of the training windows. Hyperparameters are selected on the within-subject validation split (seed~0), and the checkpoint of each run is the one with the best validation macro-F1. The test split is evaluated once per run. We report the mean and standard deviation over three seeds (within-subject) and over the five folds (cross-subject). Chance level for macro-F1 and accuracy is about 1/6.

\subsection{Models}
\paragraph{Sequence encoders:} Telemetry, EEG, PPG, and EEG+PPG are encoded by a temporal convolutional network with four blocks (64 channels, kernel size 7, stride 2 after the first block, batch normalization, ReLU), followed by mean and max pooling and a 128-unit embedding. We also evaluate \emph{mean-removed} (``m.-r.'') physiological inputs, in which each channel's mean is subtracted per window. This removes recording-specific offsets, such as the PPG signal level that depends on sensor contact, and forces the model to use signal dynamics.
\paragraph{Video encoder:} A 3D CNN with four convolutional blocks (16, 32, 64, 64 filters; $3^3$ kernels; batch normalization), max pooling, and global average pooling produces a 128-unit embedding from 25-frame clips (2.5~fps) of $48\times48$ grayscale frames, with random horizontal flips during training.
\paragraph{Training:} Each encoder has a classification head (64 units) and is trained with AdamW (weight decay $10^{-4}$), class-weighted cross-entropy, gradient clipping at 1.0, learning-rate halving on validation plateaus, and early stopping on validation macro-F1. Searched values: learning rate $\{10^{-3}, 3\cdot10^{-4}\}$ and dropout $\{0.2, 0.5\}$ for the sequence encoders; learning rate $\{10^{-3}, 3\cdot10^{-4}\}$ for the video encoder.
\paragraph{Fusion:} The encoders trained in each run are frozen; their embeddings are z-scored and concatenated, and a fusion head (256 and 128 units) is trained on the same split, with dropout, modality dropout ($\{0, 0.2\}$), and learning rate selected on the validation split. A missing modality (two trials without video) is represented by a zero embedding. As a training-free reference, we also report late fusion by averaging the encoders' class probabilities.

\subsection{Results}
\label{sec:results}
\begin{figure}[t]
  \centering
  \includegraphics[width=\columnwidth]{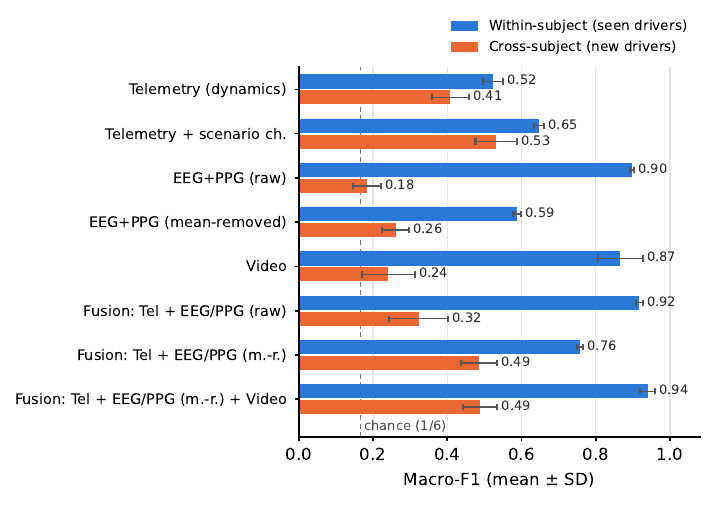}
  \caption{Macro-F1 for drivers seen during training (within-subject) and unseen drivers (cross-subject). Fusion of driving dynamics with mean-removed EEG/PPG generalizes best to new drivers.}
  \label{fig:withincross}
\end{figure}

Table~\ref{tab:results} reports all configurations and Figure~\ref{fig:withincross} contrasts the two protocols for the main ones.

\begin{table*}[t]
\caption{Six-class results (mean $\pm$ SD; within-subject: 3 seeds, cross-subject: 5 participant-disjoint folds). ``Scenario'' = traffic density, sun altitude, ambient light, snow; ``m.-r.'' = per-window mean removed. Bold: main results without scenario channels or recording offsets.}
\label{tab:results}
\small
\begin{tabular}{llcccc}
\toprule
 & & \multicolumn{2}{c}{Within-subject} & \multicolumn{2}{c}{Cross-subject (unseen drivers)} \\
\cmidrule(lr){3-4}\cmidrule(lr){5-6}
Group & Input & Acc.\ (\%) & Macro-F1 & Acc.\ (\%) & Macro-F1 \\
\midrule
\multirow{3}{*}{Telemetry} & Dynamics & 52.1 $\pm$ 2.6 & 0.524 $\pm$ 0.027 & 40.8 $\pm$ 5.3 & 0.408 $\pm$ 0.050 \\
 & Dynamics + scenario & 64.9 $\pm$ 1.4 & 0.648 $\pm$ 0.013 & 54.0 $\pm$ 7.8 & 0.532 $\pm$ 0.056 \\
 & Dynamics + scenario + position & 66.3 $\pm$ 0.9 & 0.661 $\pm$ 0.009 & 52.3 $\pm$ 6.9 & 0.519 $\pm$ 0.049 \\
\midrule
\multirow{6}{*}{Physiology} & EEG & 69.3 $\pm$ 1.8 & 0.692 $\pm$ 0.017 & 25.9 $\pm$ 6.0 & 0.253 $\pm$ 0.060 \\
 & PPG & 80.7 $\pm$ 0.2 & 0.804 $\pm$ 0.001 & 15.7 $\pm$ 1.4 & 0.140 $\pm$ 0.029 \\
 & EEG + PPG & 89.7 $\pm$ 0.5 & 0.898 $\pm$ 0.005 & 19.6 $\pm$ 4.6 & 0.184 $\pm$ 0.037 \\
 & EEG (m.-r.) & 60.3 $\pm$ 2.2 & 0.601 $\pm$ 0.022 & 28.7 $\pm$ 4.6 & 0.266 $\pm$ 0.044 \\
 & PPG (m.-r.) & 47.0 $\pm$ 1.4 & 0.455 $\pm$ 0.009 & 31.2 $\pm$ 3.8 & 0.276 $\pm$ 0.044 \\
 & EEG + PPG (m.-r.) & 59.5 $\pm$ 0.9 & 0.587 $\pm$ 0.012 & 29.2 $\pm$ 3.0 & 0.261 $\pm$ 0.036 \\
\midrule
Video & $48\times48$ grayscale & 86.7 $\pm$ 5.9 & 0.866 $\pm$ 0.060 & 30.8 $\pm$ 4.5 & 0.241 $\pm$ 0.071 \\
\midrule
\multirow{7}{*}{Fusion} & Telemetry + EEG/PPG & 91.6 $\pm$ 1.0 & 0.916 $\pm$ 0.009 & 34.1 $\pm$ 7.6 & 0.323 $\pm$ 0.079 \\
 & \textbf{Telemetry + EEG/PPG (m.-r.)} & \textbf{76.0 $\pm$ 0.8} & \textbf{0.758 $\pm$ 0.008} & \textbf{49.3 $\pm$ 5.1} & \textbf{0.485 $\pm$ 0.048} \\
 & Telemetry + video & 91.5 $\pm$ 2.3 & 0.915 $\pm$ 0.023 & 46.5 $\pm$ 4.9 & 0.444 $\pm$ 0.041 \\
 & EEG/PPG + video & 97.9 $\pm$ 0.7 & 0.979 $\pm$ 0.008 & 34.7 $\pm$ 6.2 & 0.312 $\pm$ 0.060 \\
 & Telemetry + EEG/PPG + video & 97.7 $\pm$ 0.4 & 0.977 $\pm$ 0.004 & 42.4 $\pm$ 11.1 & 0.392 $\pm$ 0.115 \\
 & \textbf{Telemetry + EEG/PPG (m.-r.) + video} & \textbf{94.0 $\pm$ 1.9} & \textbf{0.939 $\pm$ 0.020} & \textbf{50.5 $\pm$ 5.0} & \textbf{0.489 $\pm$ 0.046} \\
 & Telemetry + scenario + EEG/PPG + video & 97.7 $\pm$ 0.4 & 0.978 $\pm$ 0.004 & 48.5 $\pm$ 9.4 & 0.441 $\pm$ 0.092 \\
\bottomrule
\end{tabular}
\end{table*}

\paragraph{Seen vs.\ unseen drivers:}
Physiology and video achieve high within-subject scores (EEG+PPG 89.7\%, video 86.7\%), and both benefit from information that is specific to an individual and a session, such as the PPG signal level, which depends on sensor contact, and the driver's appearance. Such individual characteristics are valuable for personalized monitoring, but they do not carry over to new drivers (EEG+PPG 19.6\%, video 30.8\% cross-subject). Removing the per-window mean from the physiological signals makes the models rely on signal dynamics, which generalize better (PPG from 15.7\% to 31.2\% cross-subject).

\paragraph{Scenario context:}
Adding traffic density, sun altitude, ambient light, and snow raises telemetry from 0.408 to 0.532 cross-subject macro-F1, better in all five folds. Since these channels describe the scenario that accompanies each induced state (e.g., \textit{Drowsy} is driven at night), we report the main results without them, so that the models are evaluated on the driver's own signals; the context channels remain available for context-aware modelling.

\paragraph{Fusion improves recognition for unseen drivers:}
Without scenario channels and recording offsets, fusing driving dynamics with mean-removed EEG/PPG raises cross-subject macro-F1 from 0.408 (telemetry alone) to 0.485, and is better in all five folds (mean gain 7.7 points); within subject, it exceeds the best single input (m.-r.\ EEG, 0.601) by 15.7 points. Adding video brings no further cross-subject gain (0.489; better in two of five folds). Fusion with non-normalized physiology reaches the highest within-subject scores (0.916) but transfers less well (0.323), which underlines the value of per-window normalization for cross-driver models. A learned fusion head outperforms probability averaging (late fusion: 0.447 cross-subject for telemetry + m.-r.\ EEG/PPG).

\paragraph{Per-class behaviour:}
Table~\ref{tab:classwise} gives per-class results of the main model on unseen drivers, pooled over folds. \textit{Drowsy} is recognized best (F1 0.70), followed by \textit{Alerted} and \textit{Stressed} (0.58). The low-arousal, low-demand conditions \textit{Normal} and \textit{Relaxed} are the most similar pair (0.33 and 0.28) and are mainly confused with each other, which reflects their close physiological and behavioural profiles.

\begin{table}[t]
\caption{Per-class results of telemetry + EEG/PPG (m.-r.) + video on unseen drivers (5 folds pooled; 8{,}047 windows).}
\label{tab:classwise}
\small
\begin{tabular}{lrrrr}
\toprule
Class & Precision & Recall & F1 & Windows \\
\midrule
Alerted    & 0.58 & 0.58 & 0.58 & 1{,}277 \\
Distracted & 0.54 & 0.50 & 0.52 & 1{,}361 \\
Drowsy     & 0.67 & 0.74 & 0.70 & 1{,}451 \\
Normal     & 0.34 & 0.32 & 0.33 & 1{,}335 \\
Relaxed    & 0.33 & 0.25 & 0.28 & 1{,}360 \\
Stressed   & 0.51 & 0.67 & 0.58 & 1{,}263 \\
\bottomrule
\end{tabular}
\end{table}

\paragraph{Implications:}
For evaluation on \dataset, we recommend reporting both protocols, reporting scenario-context channels separately, and normalizing physiological inputs per window. The gap between within- and cross-subject scores (0.758 vs.\ 0.485 for the main model) quantifies the inter-driver heterogeneity that personalization and federated learning~\cite{mcmahan2017fedavg, li2020federated} aim to close; with one client per participant, temporal misalignment and client latency become additional factors~\cite{tziampazis2025oracle}.

\section{Limitations and Future Work}
\label{sec:limitations}
The \dataset recordings were made in a driving simulator with a consumer-grade four-channel EEG headband, a setup that favours analyses of slow physiological dynamics; the per-trial metadata documents the effective sampling rates. The benchmark uses compact models, including low-resolution video clips, so that it can be reproduced on standard hardware; face-cropped, higher-resolution video models are a natural extension. Future work includes extending the cohort, full-rate physiological recording, leave-one-subject-out and federated evaluation, and on-road data collection.

\section{Conclusion}
We presented \dataset, a four-modality driver mental-state dataset with telemetry, EEG, PPG, and in-cabin video across six induced states from 23 participants, released with corrected labels, original timestamps, and per-trial metadata. The benchmark shows that multimodal fusion reaches a macro-F1 of 0.94 for seen drivers and that fusing driving dynamics with normalized physiology consistently improves recognition for unseen drivers. The remaining gap between seen and unseen drivers makes \dataset a concrete testbed for personalized and federated driver monitoring. 
\section*{Data and Code Availability}
\dataset is archived in DaRUS, the research data repository of the University of Stuttgart, under \href{https://doi.org/\datasetdoi}{doi:\datasetdoi}. 
Raw logs and video are available under a data-use agreement (DUA); metadata, documentation, and the access procedure are public on the DaRUS landing page. The release includes the conversion script (\texttt{fix\_\allowbreak dataset.py}), the benchmark script (\texttt{train\_\allowbreak benchmark.py}) with the exact configuration used here, the per-run results, and the acquisition software.

\section*{Ethics and Responsible Use}
\dataset contains EEG, PPG, and cabin-facing video, which constitute special categories of personal data under the GDPR\@. Released files contain no direct identifiers; participants are identified only by pseudonymous IDs, and the key linking IDs to identities is not part of the release. Access to raw physiological logs and video is gated through a DUA that prohibits re-identification attempts, unapproved biometric profiling, and use of derived models in non-consented surveillance, including workplace monitoring and insurance scoring. We encourage users to report participant-wise performance before drawing conclusions about deployment readiness.

\bibliographystyle{ACM-Reference-Format}
\bibliography{references}

\end{document}